\documentclass[aps,prd,onecolumn,groupedaddress,showpacs,nofootinbib,amssymb]{revtex4}
\usepackage[T1]{fontenc}
\usepackage[latin1]{inputenc}
\usepackage{graphicx}
\usepackage[english]{babel}
\usepackage{amsmath}
\usepackage{amssymb}
\usepackage{amsfonts}

\begin{document}

\title{$AdS_{2} \times S^{2}$ geometries and the extreme 
quantum-corrected black holes}

\author{Jerzy Matyjasek$^1$, Dariusz Tryniecki$^2$}

\affiliation{\it $^1$Department of Theoretical Physics, 
Maria Curie-Sk\l odowska University, \\
pl. Marii Sk\l odowskiej 1, 20-031 Lublin, Poland\\
email: jurek@kft.umcs.lublin.pl\\
$^2$Institute of Theoretical Physics, Wroc\l aw University,\\
pl. Maxa Borna 1, 20-504 Wroc\l aw, Poland.}

\begin{abstract}
The second-order term of the approximate stress-energy tensor of the
quantized massive scalar field in the Bertotti-Robinson and Reissner-
Nordstr\"om spacetimes is constructed within the framework of the
Schwinger-DeWitt method. It is shown that although the Bertotti-
Robinson geometry is a self-consistent solution of the ($\Lambda =0$)
semiclassical Einstein field equations with the source term given by
the leading term of the renormalized stress-energy tensor, it does not
remain so when the next-to-leading term is taken into account and
requires the introduction of a cosmological term. The addition of the
electric charge to the system does not change this behavior. The near
horizon geometry of the extreme quantum-corrected Reissner-Nordstr\"om
black hole is analyzed. It has the $AdS_{2} \times S^{2}$ topology and
the sum of the  curvature radii of the two dimensional submanifolds is
proportional to the trace of the second order term. It suggests  that
the ``minimal'' approximation  should be constructed from the first
two terms of the Schwinger-DeWitt expansion

\end{abstract}


\maketitle

\section{Introduction}
\label{intro}

According to our present understanding the physical content of the
quantum field theory formulated in a general curved background is
encoded in its most important observable -- the renormalized stress-
energy tensor, $T^{ab},$ calculated in the physically motivated state
or states~\cite{pcwd}. Although interesting in its own, the stress-
energy tensor plays a crucial role in various applications, most
important of which is its role played in the semiclassical Einstein
field equations. Indeed, treating the stress-energy tensor as a source
term it is possible, in principle, to investigate the evolution of the
system in a self-consistent way unless the quantum gravity effects
become dominant. Unfortunately, this program is hard to execute as it
requires a detailed knowledge of the functional dependence of  $T^{ab}
$ on the metric tensor or at least on a wide class of metrics.
Moreover, the semi-classical field equations comprise a complicated
system of the nonlinear partial differential equations, which
generally cannot be solved analytically. It is natural, therefore,
that to address -- at least partially -- this problem, one should
refer either to approximations or to numerical methods. Quite popular
are also the hybrid methods, in which the ``exact'' results are
divided into the analytic part and (presumably) small numerical
correction.

It seems that for the massive fields in a large mass limit, i.e., when
the Compton length, $\lambda_{C},$ is much smaller than the
characteristic radius of curvature, ${\mathbb L},$ (where the latter
means any characteristic length scale of the spacetime), the
approximation based on the asymptotic Schwinger-DeWitt expansion is
sufficiently general~\cite {Julian,Bryce1,Bryce2}. Since the nonlocal
contribution to the effective action can be neglected it is expected
that the method yields reasonable results provided the curvature of
the background geometry is well above the Planck regime and its
temporal changes remain small. In spite of the above restrictions
there is still a wide class of geometries in which this approximation
can be successfully applied. At the moment we know the $m^{-2}$ part
of the renormalized stress-energy tensor for scalar, spinor and vector
field~\cite{kocio1,kocio2} whereas constructed only recently the next-
to-leading~\cite{kocio3} and next-to-next-to-leading terms are known
for the scalar field with an arbitrary curvature coupling.

For a neutral massive scalar field satisfying the generalized Klein-
Gordon equation
\begin{equation}
     \left( \Box \,-\,\xi R\,-\,m^{2}\right) \phi \,=\,0,  
	                                          \label{wave}
\end{equation}
where $\xi $ is the coupling parameter, the Schwinger-DeWitt expansion
of the renormalized approximate effective action, $W_{R},$ assumes the
form \cite{fz1,fz2,fz3}
\begin{equation}
W_{R}\,=\,\frac{1}{32\pi^{2}}\sum_{n=3}^{\infty}
\frac{(n-3)!}{(m^{2})^{n-2}}\int d^{4}x \sqrt{g} [a_{n}],
                           \label{Weff}
\end{equation}
where $[a_{n}]$ is the coincidence limit of the $n-$th Hadamard-
Minakshisundaram-DeWitt coefficient $a_{n}(x,x'), $ which has the
dimensionality of $[length]^{2n}$ and is constructed solely from the
curvature tensor, its covariant derivatives up to $2n -2$ order, and
appropriate contractions. The spin of the field enters the effective
action only through the constant numerical coefficients.

Recently, there has been a renewal of interest in calculations of
various important bi-tensors and the objects constructed form them,
with the special emphasis put on the coincidence limit of the
coefficients $a_{n}(x,x')$ and the renormalized stress-energy
tensor~\cite{kocio1,kocio2,lemos,kocio3,Dec1,Ott}. This has been
largely stimulated by new methods of computations and rapid
development of the computer algebra systems. Moreover, due to
increasing power of home computers the complexity of calculations that
can be  successfully executed increased enormously and recent findings
substantially extended previous results
\cite{Bryce1,sakai,gi11,gi12,Parker,AvraPhD,Avraros,Avranuc,Amsterdamski,Ven}.

Having at one's disposal the approximation of the renormalized
effective action, the stress-energy tensor can be evaluated by means
of the standard formula
\begin{equation}
    \frac{2}{\sqrt{g}}\frac{\delta }{\delta g_{ab}}W_{R}
    = T^{ab}.
                            \label{def1}
\end{equation}
As the complexity of $[a_{n}]$  grows rapidly with
increasing $n,$ a practical use of the effective action (\ref{Weff})
is limited either to special geometries or the first few coefficients. 

Recently, we have calculated the next-to-leading, $T_{i}^{(2)j},$ and
the next-to-next-to-leading,  $T_{i}^{(3)j},$  term of the approximate
stress-energy tensor of the quantized massive scalar field with an
arbitrary curvature coupling. (The first order approximation of the
stress-energy tensor of the quantized massive scalar, spinor and
vector fields  has been constructed in Refs.~5,6.)
%%%%%
%%% Refs.~\cite{kocio1,kocio2}). 
%%%%%
Since the full results expressed in terms of the Riemann tensor are
far too complicated to be displayed in the text, we have presented the
simplified form of $T_{i}^{(2)j}$ for the static and spherically-
symmetric geometries in Ref.~7, to which the interested reader is
referred for the formulas and technical details.
%%%%%
%%% Ref.~\cite{kocio3} 
%%%%%
The calculations of the coefficients $[a_{4}]$ and $[a_{5}]$ and their
functional derivatives with respect to the metric  tensor have been
carried out with the aid of FORM~\cite{Vermaseren} and its multithread
version TFORM~\cite{Vermaseren-Tentyukov}, and are stored in the FORM
format. Here we shall concentrate  on the simplest, yet
calculationally involved case, in which the approximate effective
action can be written as
\begin{equation}
     W_{R}=\frac{1}{32\pi ^{2}}\int d^{4}x\sqrt{g}\left( \frac{1}{m^{2}}
	  \left[ a_{3}\right]+ \frac{1}{m^{4}}\left[a_{4}\right]\right).  
	                                               \label{Weff1}
\end{equation}
The next-to-next-to-leading term constructed from $[a_{5}]$ will be
briefly discussed in the context of the Bertotti-Robinson geometry in
Sec.~\ref{drugi}

The reasons for carrying out the higher-order calculations - besides
natural curiosity - are numerous. First, it should be emphasized that
although the Schwinger-DeWitt expansion is used, it does not mean that
the second order term is negligible. Indeed, it may be a considerable
fraction of the first order term, leading to improvement of the
approximation and the stress-energy tensor of the massive field in the
Schwarzschild spacetime is an interesting example of such
behavior~\cite{kocio3}. Preliminary calculations reveal similar
behavior in the Reissner-Nordstr\"om geometry. Moreover, the higher
order terms may dramatically change the type of the solutions of the
semiclassical Einstein field equations. An interesting example in this
regard, as we shall see, is given by the Bertotti-Robinson
geometry~\cite{robinson,bertotti}. It can be shown that although the
Bertotti-Robinson geometry is a self-consistent solution of the
semiclassical Einstein field equations with the source term given
solely by the leading term of the renormalized stress-energy
tensor~\cite{Kofman1,Kofman2,kocio1,Olek} it does not remain so when
the next-to-leading term is taken into account. To guarantee that the
Bertotti-Robinson spacetime is the solution of the semiclassical
equations one has to introduce the (negative) cosmological constant.
It should be noted that addition of the electric charge to the system
does not change this behavior. It suggests that the first two terms of
the stress-energy tensor constructed within the framework of the
Schwinger-DeWitt method comprise, in a sense, a minimal approximation.

In this note we shall analyze the semiclassical Einstein field
equations with the stress-energy tensor given by the first two terms 
obtained within the framework of the Schwinger-DeWitt method for 
the Bertotti-Robinson spacetime and its deformations to general 
$AdS_{2}\times S^{2}$ geometries with and without an electric charge. 
(For a different source term see Ref.~30 and the references therein.)
%%%%%%%%%%
%%%%% Ref.~\cite{Solo}. 
%%%%%%%%%%
As is well known the solutions with the maximally symmetric two-dimensional 
subspaces universally describe the near-horizon geometry of the extreme black
holes~\cite{carter,ginsparg,Paul1,MannRo,Hawk,CVZ,Dadhich,lemos1,lemos0,jaExtr,Klimek}.
%%%%%%%%%
One expects, therefore, that it is possible to analyze the vicinity 
of the degenerate horizon without detailed knowledge of the
renormalized stress-energy tensor. To show that this is indeed the case
we shall calculate the second-order
approximation of the renormalized stress-energy tensor in the
Reissner-Nordstr\"om geometry, and, solving the semiclassical Einstein
field equations, we shall construct its quantum corrected counterpart.
The vicinity of the event horizon of the extreme limit is, as
expected, precisely described by the relations found for the  $AdS_{2}
\times S^{2}$ geometries and it is the next-to-leading term that
deforms the Bertotti-Robinson solution. This behavior can be
attributed to a nonvanishing trace of $T_{i}^{(2)j}$ tensor in the
Bertotti-Robnson spacetime. Throughout the paper the geometric system
of units is used and our conventions follow the conventions of MTW.
%%%%%%%%%%%%%%%%
%%%%%%%%%%%%%%%%
%% NOWA ORGANIZACJA%%
\section{$AdS_{2} \times S^{2}$ geometries as the self-consistent 
semiclassical solutions}
\label{drugi}

Now, let us confine ourselves to the geometries satisfying 
simplifying assumptions: homogeneity ($R_{abcd;e}=0$), vanishing of
the Weyl tensor ($C_{abcd} =0$) and $R=0.$ Since the Weyl tensor and
the curvature scalar vanish, the stress-energy tensor can be expressed
solely in terms of the Ricci tensor. However, it should be noted that 
if all three conditions are satisfied the Ricci tensor obeys additionally
the Rainich condition
\begin{equation}
R_{i}^{a} R_{a}^{j} = \frac{1}{4}R_{a b} R^{a b} \delta_{i}^{j}.
\end{equation}
Making use of the general
formulas presented in Refs. 5 and 6, the first order approximation to
the stress-energy tensor  can be written in the form 
\begin{eqnarray}
32\pi^{2} m^{2} T^{(1) i j} &=&
-\frac{5 - 14 \xi}{1260} R_{a b}R^{a b} R^{i j} .
\label{tep1}
\end{eqnarray}
On the other hand, the general form of the  next-to-leading term
constructed from  the coefficient $[a_{4}]$ is extremely complicated,
nevertheless, when the above conditions are satisfied, it simplifies
considerably and reads
\begin{eqnarray}
32\pi^{2} m^{2}  T^{(2)i j}&=& 
-\frac{11}{75600} R_{a b} R^{a b} R_{c d}R^{c d} g^{i j}.
\label{tep2}
\end{eqnarray}
Although the stress-energy tensor as given by Eqs.~(\ref{tep1}) 
and (\ref{tep2}) has been constructed for  a special class of geometries
it can be used, as we shall see, in a more general context.
%%%%%%
%%%%%%
%% Przeredagowane
%%

To proceed further let us consider a slightly more general line
element of the form
\begin{equation}
ds^{2} = \frac{1}{h y^{2}}\left( -dt^{2} + 
dy^{2}\right) + r_{0}^{2} d\Omega^{2},
                                                  \label{br_gen}
\end{equation}
for which each nonvanishing component of the Weyl tensor contains the
factor $h r_{0}^{2} -1.$ Topologically it is $ AdS_{2} \times  S^2,$
i.e., a direct product of the two-dimensional anti-de Sitter geometry
and two sphere. The curvature scalar of this geometry is a sum of the
curvatures of the subspaces
\begin{equation}
R =K_{{\rm AdS}_{2}} + K_{ {\rm S}^2},
\end{equation}
where $K_{{\rm AdS}_{2}}=-2 h$ and $K_{ {\rm S}^2}=2/r_{0}^{2}.$ For
$h=1/r_{0}^{2}$ both the Weyl tensor and the curvature scalar vanish
and we shall refer to such solutions as Bertotti-Robinson geometry
~\cite{robinson,bertotti}.

Other useful representations of (\ref{br_gen}) can be obtained through 
the coordinate transformations 
\begin{equation}
h^{1/2} t = e^{\tau} \coth \chi, \hspace{1cm} h^{1/2} 
y = e^{\tau} \sinh^{-1} \chi
\end{equation}
and
\begin{equation}
\sinh^{2}\chi ={\cal R} h -1, \hspace{1cm}  \tau h = T
\end{equation}
and are given by
\begin{equation}
ds^{2} = \frac{1}{h}\left( -\sinh^{2} \chi dt^{2} 
+ d\chi^{2}\right) + r_{0}^{2} d\Omega^{2}
                        \label{hiperbol}
\end{equation}
and
\begin{equation}
ds^{2} = -\left( {\cal R}^{2} h -1\right) dT^{2} 
+ \frac{d{\cal R}^{2}}{\left({\cal R}^{2} h -1 \right)} 
+ r_{0}^{2} d\Omega^{2},
\end{equation}
respectively.

Making use in Eqs.~(\ref{tep1}) and (\ref{tep2}) of the explicit form 
of the Ricci tensor of the Bertotti-Robinson line element 
\begin{equation}
R_{i}^{j} =\frac{1}{r_{0}^{2}} {\rm diag}[-1,-1,1,1],
\end{equation}
one has
\begin{equation}
8\pi T_{i}^{(1)j} =-\frac{5-14\xi}{1260\pi m^{2}r_{0}^{6}} 
{\rm diag}[-1,-1,1,1]
          \label{first}
\end{equation}
and
\begin{equation}
8 \pi T_{i}^{(2)j} =-\frac{11}{18900\pi m^{4}r_{0}^{8}} 
{\rm diag}[1,1,1,1].
          \label{second} 
\end{equation}
%%%
Although the higher order terms $(n \geq 4)$ of the stress-energy
tensor are not known, one expects that there will be massive
simplifications for the Bertotti-Robinson class of metrics. Indeed,
dimensional considerations suggest that the higher-order terms of the
stress-energy tensor can be divided into two generic groups based on
their trace. The ``odd'' tensors
\begin{equation}
T_{i}^{(2n-1)j} = \frac{A^{(2n-1)}}{\pi^{2}(m^{2})^{2n-1}r_{0}^{4 n+2}} 
{\rm diag}[-1, -1, 1, 1]
\end{equation}
are traceless, whereas the ``even'' tensors 
\begin{equation}
T_{i}^{(2n)j} = \frac{A^{(2n)}}{\pi^{2}(m^{2})^{2n}r_{0}^{4 n+4}} 
{\rm diag}[1, 1, 1, 1]
\end{equation}
are characterized by a nonzero trace. The $n-$th tensor, $T_{i}^{(n)
j},$ is constructed from the integrated coincidence limit $[a_{n+2}]$
and $A^{k}$ are  the parameters depending on the spin of the field and
(possibly) on the type of the curvature coupling. For example, the
tensor $T_{i}^{(3)j}$ of the massive scalar field constructed from 
the coefficient $[a_{5}]$ is given by
\begin{equation}
8\pi T_{i}^{(3)j} =-\frac{179-242\xi}{103950\pi m^{6} r_{0}^{10}}
{\rm diag}[-1,-1,1,1].
\end{equation}

The particular form of the stress-energy tensor suggests that the
Bertotti-Robinson geometry can be, in certain situations, a self-
consistent solution of the semiclassical Einstein field equations with
or without classical electromagnetic field. Indeed, it can be shown
that although the Bertotti-Robinson geometry for $\xi > 5/14$ is the
exact solution of the $(\Lambda =0)$ semiclassical Einstein field
equations with the source term given solely by the renormalized
stress- energy tensor $T^{(1)j}_{i},$ it ceases to be so when the
next-to- leading term is included. Inclusion of the electromagnetic
field described by the stress-energy tensor of the form
\begin{equation}
8\pi T^{(em)j}_{i} = \frac{Q^{2}}{r_{0}^{4}}{\rm diag}[-1,-1,1,1] 
\end{equation}
as well as the higher terms of the renormalized stress-energy tensor
does not change this behavior. On the other hand, one expects that the
cosmological term, $\Lambda \delta_{i}^{j},$ may compensate the even
terms of the stress-energy tensor. Therefore the sum of the first two
terms of the renormalized stress-energy tensor constitutes the main
approximation and the higher order terms can be regarded as
perturbations. 

Now, let us analyze this problem more closely: The Einstein tensor in
the Bertotti-Robinson geometry is given by 
\begin{equation}
G_{i}^{j} = \frac{1}{r^{2}_{0}}{\rm diag}[-1,-1,1,1 ]
                                         \label{Ricc}
\end{equation}
and it is evident that although the Bertotti-Robinson line element is
a solution of the semiclassical Einstein field equations of the form
\begin{equation}
R_{i}^{j} -\frac{1}{2}R \delta_{i}^{j} = 8\pi T_{i}^{(1)j},
                                      \label{semiB}
\end{equation}
the addition of the next-to-leading term requires the presence of the
cosmological constant in the equations:
\begin{equation}
R_{i}^{j} -\frac{1}{2}R \delta_{i}^{j} + \Lambda \delta_{i}^{j}  = 
8\pi \left( T_{i}^{(em)j} + T_{i}^{(1)j} +T_{i}^{(2)j} \right),
                                      \label{semiL}
\end{equation}
where, for generality, the stress-energy tensor of the electromagnetic
field has been added. It is simply because the trace of
Eq.~(\ref{semiL}) with $\Lambda =0$ leads to a contradiction. The two
independent equations, therefore, are
\begin{equation}
-\frac{Q^{2}}{r_{0}^{4}} +\frac{5-14 \xi}{1260 \pi m^{2} r_{0}^{6}} 
- \frac{11}{18900 \pi m^{4} r_{0}^{8}}
+\frac{1}{r_{0}^{2}} -\Lambda =0 
                              \label{BRt}
\end{equation}
and
\begin{equation}
\frac{Q^{2}}{r_{0}^{4}}-\frac{5-14 \xi}{1260 \pi m^{2} r_{0}^{6}}
- \frac{11}{18900 \pi m^{4} r_{0}^{8}}
-\frac{1}{r_{0}^{2}} -\Lambda =0 ,
                             \label{BRr}
\end{equation}
and this system can easily be solved with respect to $r_{0}$ and
$\Lambda.$ Simple manipulations give
\begin{equation}
r_{0} =\sqrt{ \frac{Q^{2}-\sqrt{ Q^{4} -4 \beta}}{2} },
\hspace{1cm} \Lambda =\frac{-16 \gamma}{\left( Q^{2} 
-\sqrt{Q^{4}-4 \beta}  \right)^{4}}
\end{equation}
and
\begin{equation}
r_{0} =\sqrt{ \frac{Q^{2}+\sqrt{ Q^{4} -4 \beta}}{2} }, 
\hspace{1cm} \Lambda =\frac{-16 \gamma}{\left( Q^{2} 
+\sqrt{Q^{4}-4 \beta}  \right)^{4}},
\end{equation}
where
\begin{equation}
\beta = \frac{5-14\xi}{1260 \pi m^{2}},  \hspace{10mm} 
\gamma = \frac{11}{18900\pi m^{4}}.
\end{equation}
For purely quantum source, i.e., when $Q=0$ one has
\begin{equation}
r_{0} =\left( \frac{14\xi -5}{1260 \pi m^{2}} \right)^{1/4}, \hspace{1cm}
\Lambda = -\frac{11}{18900\pi m^{4}r_{0}^{8}} = -\frac{924 \pi}{\left(5-14\xi\right)^{2}}
\end{equation}
and the self-consistent solution is possible for $\xi >5/14.$ It
should be noted that neither the minimal coupling nor the conformal
one belongs to the admissible range of $\xi.$   On the other hand, for
$\beta \ll Q^{4},$ one has
\begin{equation}
r_{0} = |Q| -\frac{\beta}{2 |Q|^{3}}, \hspace{1cm} \Lambda = -\frac{8\gamma}{Q^{8}}.
\end{equation}
Note, that the condition of applicability of the Schwinger-DeWitt
method $\lambda_{C}/L \ll 1$ now reduces to  $m r_{0} \gg 1.$

Before going further let us consider a general $AdS_{2} \times S^{2}$
geometry as a solution of the Einstein equations with $\Lambda=0$ and
$Q\neq 0.$ The foregoing discussion shows that quantum processes
deform the Bertotti-Robinson spacetime. The stress-energy tensor
$T_{i}^{(2)j}$ of the massive scalar field in a general $AdS_{2}
\times S_{2}$ geometry is rather complicated. Fortunately, since we
are interested in small deformations of the Bertotti-Robinson
geometry, to construct the approximate solution it is sufficient to
know  the tensors (\ref{first}) and (\ref{second}) as the only
modification of the equations comes from the Einstein tensor. Indeed,
putting in the line element~(\ref{hiperbol}) 
$\rho^{2} = r_{0}^{2} + \kappa,$ where $\rho^{2} =1/h$ 
and $\kappa$ measures the deformation,
one concludes that  Eq.~(\ref{BRt}) remains unchanged, whereas the
radial component acquires a small correction:
\begin{equation}
\frac{\kappa}{r_{0}^{4}} + \frac{Q^{2}}{r_{0}^{4}}
-\frac{5-14 \xi}{1260 \pi m^{2} r_{0}^{6}}
- \frac{11}{18900 \pi m^{4} r_{0}^{8}}
-\frac{1}{r_{0}^{2}} =0 .
                             \label{GBRr}
\end{equation}
Solving the system (\ref{BRt}) and (\ref{GBRr}), one obtains
\begin{equation}
\kappa = \frac{11}{9450\pi m^{4} r_{0}^{4}}
                   \label{kappa}
\end{equation} 
and the equation relating $Q$ and $r_{0}$:
\begin{equation}
r_{0}^{2} = Q^{2} - \frac{4-21\eta}{1890\pi m^{2} Q^{2}} +
\frac{11}{18900 \pi m^{4} Q^4}.
                     \label{relation_extr}
\end{equation}
On general grounds one expects that the relations (\ref{kappa}) and
(\ref{relation_extr}) describe the near horizon geometry of the
quantum-corrected extreme Reissner-Nordstr\"om black hole. In the next
section we shall show that this assertion is true.

\section{The near horizon geometry of the extreme quantum corrected
Reissner-Nordstr\"om black hole}
\label{trzeci}

The results of the foregoing analysis can be related to the near-
horizon geometry of the extreme black hole being the solution of the
semiclassical Einstein field equations with the source term given by
the sum of the stress-energy tensor of the electromagnetic field and
the quantum part describing the contribution of the quantized field.
The latter is constructed form the first two terms of the
series~(\ref{Weff}). Indeed, employing the general formulas describing
the stress-energy of the quantized massive fields presented
in Ref.~7
%%%%%
%%% Ref.~\cite{kocio3} 
%%%%%
one can solve the semiclassical equations, treating the nonclassical
part of the source term as a small perturbation. The results can be
expressed in terms of the exact position of the event horizon,
$r_{+},$ and the total charge, $Q.$ To simplify the calculations,
especially to keep control of the order of terms in complicated series
expansions, we shall introduce a dimensionless parameter,
$\varepsilon,$ substituting $T_{i}^{(k)j} \to \varepsilon T_{i}^{(k)
j}.$ We shall put $\varepsilon =1$ in the final stage of the
calculations.

The general static and spherically symmetric metric can be written in
the form
\begin{equation}
ds^{2} = -e^{2\psi(r)} f(r) dt^{2} + \frac{1}{f(r)}dr^{2} + r^{2}
\left( d\theta^{2} + \sin^{2} \theta d\phi^{2}\right),
                                          \label{ss}
\end{equation}
where $f(r)$ and $\psi(r)$ are two unknown functions. With the
boundary conditions $f(r_{+}) =0$ and $\psi(\infty) =0$ the
semiclassical Einstein field equations can be solved formally to give

\begin{equation} 
    f\left( r\right)
    =1-\frac{r_{+}}{r}+\frac{Q^{2}}{r^{2}} -
    \frac{Q^{2}}{r r_{+}}+\frac{8\pi\varepsilon }{r}%
    \int_{r_{+}}^{r}dr^{\prime }(r^{\prime })^{2}
    \left( T_{t}^{(1)t}+ T_{t}^{(2)t}\right) 
             \label{lou1}  
\end{equation}  
and
\begin{equation}
    \psi \left( r\right) \,=\,
    4\pi\varepsilon \int_{\infty }^{r} dr^{\prime }
r^{\prime }\left( T_{r}^{(1)r}+T_{r}^{(2)r} - T_{t}^{(1)t}- T_{t}^{(2)t} \right)
f_{0}^{-1}(r') ,  
                                                     \label{lou2}
\end{equation}
where
\begin{equation}
f_{0}(r) = 1-\frac{r_{+}}{r}+\frac{Q^{2}}{r^{2}} -
    \frac{Q^{2}}{r r_{+}}
\end{equation}
and the tensors $T^{(k)j}_{i}$ are to be calculated for 
$f(r) = f_{0}(r)$ and $\psi(r) =0.$

Since the stress-energy tensor $T_{i}^{(1)j}$ is well known 
(see Refs.~42,5,6,43)
%%%%%
%%% Refs.~\cite{AHS95,kocio1,kocio2,Taylor,jirinek03b}) 
%%%%%
we shall not display it here. On the other hand, $T_{i}^{(2)j}$ of the
massive scalar field in the Reissner-Nordstr\"om spacetime has not
been constructed previously.
%%%%%
%%% Ref.~\cite{AHS95}
%%%%%
In Ref.~42, the renormalized stress-energy tensor has been calculated
numerically and it was shown that the first order term provides a
reasonable approximation, leaving however a room for improvement. It
is expected that such an improvement can be achieved by inclusion of
the second order term, which may constitute a considerable portion of
$T_{i}^{(1)j}.$ The general form of the tensor $T_{i}^{(2)j}$ (i.e.,
the one expressed in terms of the Riemann tensor, its covariant
derivatives and contractions) is extremely complicated. Fortunately,
there are massive simplifications for the line element describing
static and spherically-symmetric geometry. Making use of  Eqs.~8-17 of
Ref.~7
%%%%%
%%% Ref~\cite{kocio3} 
%%%%%
for the  Reissner-Nordstr\"om  line element, putting $q=Q/r_{+},$  
$x = r/r_{+}$ and $\eta =\xi -1/6,$ after some algebra, one has
\begin{equation}
T_{i}^{(2)j} = \frac{1}{32\pi m^{4}x^{10} r_{+}^{8}}\left(C_{i}^{j} 
+ \eta D_{i}^{j}\right),
\end{equation}
where 
\begin{eqnarray}
C_{t}^{t} &=&-{\frac {50369 q^{6}}{1050 x^{2}}}
-{\frac {11}{5}}
+{\frac {4089 q^{2}}{175 x}}
+{\frac {83611 q^{8}}{3150 x^{3}}}
-{\frac {13583 q^{8}}{4200 x^{2}}}
+{\frac {4089 q^{4}}{175 x}}
+{\frac {2833 q^{6}}{525 x}}
-{\frac {26 q^{2}}{15}} \nonumber \\
&+&{\frac {83611 q^{2}}{3150 x^{3}}}
-{\frac {50369 q^{2}}{1050 x^{2}}}
-{\frac {287009 q^{4}}{3150 x^{4}}}
+{\frac {579337 q^{4}}{3150 x^{3}}}
+{\frac {191678 q^{6}}{1575 x^{5}}}
+{\frac {579337 q^{6}}{3150 x^{3}}}
-{\frac {11 q^{4}}{5}}\nonumber \\
&-&{\frac {13583}{4200 x^{2}}}
+{\frac {191678 q^{8}}{1575 x^{5}}}
-{\frac {287009 q^{8}}{3150 x^{4}}}
-{\frac {48049 q^{4}}{420 x^{2}}}
-{\frac {255229 q^{8}}{4725 x^{6}}}
-{\frac {1241053 q^{6}}{4725 x^{4}}}
+{\frac {2833}{525 x}}\nonumber \\
             \label{sta}
\end{eqnarray}

\begin{eqnarray}
D_{t}^{t} &=&\frac {173777 q^{4}}{70x^{2}}
+\frac {1649 q^{8}}{28 x^{2}}
-94 \frac {q^{6}}{x}
+\frac {330616 q^{8}}{315 x^{6}}
+36 
-\frac {160207 q^{4}}{42 x^{3}}
+\frac {390577 q^{8}}{210 x^{10}}\nonumber \\
&-&\frac {160207 q^{6}}{42 x^{10}}
-\frac {71087 q^{8}}{126 x^{3}}
-\frac {36016 q^{6}}{15 x^{5}}
-\frac {36016 q^{8}}{15 x^{5}}
-\frac {19374 q^{2}}{35 x}
-\frac {19374 q^{4}}{35 x} +36 q^{4}\nonumber \\
&+&\frac {72851 q^{6}}{70 x^{2}}
+\frac {552233 q^{6}}{105 x^{4}}
-\frac {71087 q^{2}}{126 x^{3}}
+\frac {72851 q^{2}}{70 x^{2}}
+\frac {390577 q^{4}}{210 x^{4}}
+\frac {1649}{28 x^{2}} 
-\frac{94}{x}
+72 q^{2}
                                \nonumber \\
\end{eqnarray}
and
\begin{eqnarray}
C_{r}^{r}&=&
{\frac {2143 q^{6}}{150 x^{2}}}
+\frac{3}{5}
-{\frac {327 q^{2}}{35 x}}
-{\frac {19907 q^{8}}{3150 x^{3}}}
+{\frac {2753 q^{8}}{4200 x^{2}}}
-{\frac {327 q^{4}}{35 x}}
-{\frac {97 q^{6}}{75 x}}
-{\frac {97}{75 x}}
\nonumber \\
&-&{\frac {19907 q^{2}}{3150 x^{3}}}
+{\frac {2143 q^{2}}{150 x^{2}}}
+{\frac {11581 q^{4}}{630 x^{4}}}
-{\frac {431867 q^{4}}{9450 x^{3}}}
-{\frac {302 q^{6}}{15 x^{5}}}
-{\frac {431867 q^{6}}{9450 x^{3}}}
+\frac{3}{5}{q}^{4}
\nonumber \\
&+&{\frac {2753}{4200 x^{2}}}
-{\frac {302 q^{8}}{15 x^{5}}}
+{\frac {11581 q^{8}}{630 x^{4}}}
+{\frac {673339 q^{4}}{18900 x^{2}}}
+{\frac {34463 q^{8}}{4725 x^{6}}}
+{\frac {83831 q^{6}}{1575 x^{4}}}
+{\frac {166}{105}}\,{q}^{2}
                          \nonumber \\
\end{eqnarray}

\begin{eqnarray}
D_{r}^{r} &=&-{\frac {45241 q^{4}}{90 x^{2}}}
-{\frac {291 q^{8}}{28 x^{2}}}
+{\frac {151 q^{6}}{7 x}}
-{\frac {30056 q^{8}}{315 x^{6}}}
-{\frac {72 }{7}}
+{\frac {403111 q^{4}}{630 x^{3}}}
-{\frac {3487 q^{8}}{14 x^{4}}}\nonumber \\
&+&{\frac {403111 q^{6}}{630 x^{3}}}
+{\frac {1567 q^{8}}{18 x^{3}}}
+{\frac {4064 q^{6}}{15 x^{5}}}
+{\frac {4064 q^{8}}{15 x^{5}}}
+{\frac {663 q^{2}}{5 x}}
+{\frac {663 q^{4}}{5 x}}
-{\frac {42857 q^{6}}{210 x^{2}}}\nonumber \\
&-&{\frac {33197 q^{6}}{45 x^{4}}}
+{\frac {1567 q^{2}}{18 x^{3}}}
-{\frac {42857 q^{2}}{210 x^{2}}}
-{\frac {3487 q^{4}}{14 x^{4}}}
-{\frac {291}{28 x^{2}}}
+{\frac {151}{7 x}}
-{\frac {144}{7}}\,{q}^{2}
-{\frac {72}{7}}\,{q}^{4}. 
                           \nonumber \\
                \label{fin}
\end{eqnarray}
Only the time, $T_{t}^{(2)t},$ and the radial, $T_{r}^{(2)r},$ 
component of the stress-energy tensor are presented as the angular 
components $T_{\theta}^{(2)\theta}=T_{\phi}^{(2)\phi},$  can be calculated 
from the equation
$\nabla_{b} T_{a}^{b} =0,$ that, in the case on hand, gives
\begin{equation}
T^{(2)\theta}_{\theta} = T^{(2)\phi}_{\phi} = -\frac{r}{4 f_{0}}
\left(T^{(2)t}_{t} -T^{(2)r}_{r} \right)\frac{d }{dr} f_{0} +
\frac{r}{2}\frac{d }{dr}T^{(2)r}_{r}  + T^{(2)r}_{r}.
\end{equation} 
The difference $T_{r}^{(2)r} -T_{t}^{(2)t}$ factorizes as 
$f_{0}(r) P^{(2)}(r),$ where $P^{(2)}(r)$ is a polynomial in $x^{-1}.$ 
This together with the polynomial form of the components of 
$T_{i}^{j}$ guarantees the regularity of the stress-energy tensor at the event
horizon in a physical sense. Inserting the tensors $T_{i}^{(1)j}$ and
$T_{i}^{(2)j}$ into the Eqs.~(\ref{lou1}) and (\ref{lou2}) one clearly
sees that the problem of finding the line element of the quantum
corrected black hole reduces to two elementary quadratures.

The event and inner horizons of the classical Reissner-Nordstr\"om
black hole are located at $r_{+}$ and $r_{-},$ respectively. In
$(r_{+},Q)$ representation they are related by inversion with respect
to a sphere of a radius equal to the radius of the extreme black hole.
For the quantum-corrected black hole  $r_{+}$ remains unchanged
(because of the choice of the boundary conditions) whereas the inner
horizon acquires some corrections. The extremal configuration is,
therefore, determined by the condition $r_{+} =r_{-},$ which, in turn,
leads to the simple relation between $Q$ and $r_{+}$
\begin{equation}
Q^{2} = r_{+}^{2} + \varepsilon\frac{4-21\eta}{1890\pi m^{2} r_{+}^{2}} -
\varepsilon\frac{11}{18900 \pi m^{4} r_{+}^4}. 
                     \label{xtr1b}
\end{equation}
One can easily demonstrate that the equations (\ref{relation_extr})
and (\ref{xtr1b}) are equivalent. Indeed, inverting the relation
(\ref{relation_extr}), i.e., expressing the charge in terms of the
radius $r_{0}$ and subsequently putting $r_{0} = r_{+}$ one obtains
precisely Eq. (\ref{xtr1b}).

Now, to analyze the geometry of the vicinity of the degenerate event
horizon, which is our main interest here, we shall approximate the
function $f(r)$ by a parabola $\alpha (r-r_{+})(r-r_{-}),$ where
$\alpha$ is a suitably chosen parameter~\cite{Klimek}. For the near
extreme black hole one has
\begin{equation}
Q^{2} = r_{+}^{2} + \varepsilon\frac{4-21\eta}{1890\pi m^{2} r_{+}^{2}} -
\varepsilon\frac{11}{18900 \pi m^{4} r_{+}^4} -\Delta ,
\end{equation}
where $\Delta$ measures deviation from extremal configuration. The
location of the event horizon is given by $r_{+},$ whereas for $r_{-}$
one has
\begin{equation}
r_{-} = r_{+} -\Delta \left[\frac{1}{r_{+}} + \varepsilon\left(
\frac{2}{945\pi m^{2} r_{+}^{5}} - \frac{\eta}{90\pi m^{2} r_{+}^{5}} 
-\frac{11}{9450\pi m^{4}r_{+}^{7}}
\right) \right].
\end{equation}
The coefficient $\alpha$ can be calculated from the (approximate) equality
\begin{equation}
f(r_{d}) = - \alpha (r_{+} -r_{d})(r_{d} -r_{-}),
\end{equation}
where $r_{d} = (r_{+} + r_{-})/2.$
Simple calculations give
\begin{equation}
\alpha = \frac{1}{r_{+}} -\frac{11}{9450 \pi m^{4} r_{+}^{8}}\varepsilon 
+\varepsilon \Delta \left(\frac{8}{945 \pi m^{2} r_{+}^{8}} 
-\frac{2}{45 \pi m^{2} r_{+}^{8}}\eta -\frac{22}{4725\pi m^{4} r_{+}^{10}} \right).
\end{equation}
Further, making use  in the line element (\ref{ss})  of the  coordinate 
transformation 
\begin{equation}
r = r_{d} + \gamma \cosh \chi
\end{equation} 
and
\begin{equation}
t = \frac{\tau}{\Delta} e^{-\psi(r_{+})} \left[2 r_{+}^{3} + \varepsilon \left(
\frac{22}{4725\pi m^{4} r_{+}^{3}} -
\frac{4}{945 \pi m^{2} r_{+}^{8}} +\frac{1}{45 \pi m^{2} r_{+}^{8}}\eta
\right) \right],
\end{equation}
where $\gamma = r_{+} -r_{d},$  
and subsequently taking  the limit $\Delta \to 0$ one obtains
\begin{equation}
ds^{2} = \rho^{2} \left(- \sinh^{2}\chi\, d\tau^{2} + d\chi^{2} \right) + r_{+}^{2}
\left(d\theta^{2} + \sin^{2} \theta \,d\phi^{2} \right),
                     \label{AdS2}
\end{equation} 
where 
\begin{equation}
\rho^{2} = r_{+}^{2} + \varepsilon\frac{11}{9450 \pi m^{4} r_{+}^{4}}.
                  \label{dfm}
\end{equation}
This result clearly shows that the geometry of the closest vicinity of
the extreme black hole, when expanded into a whole manifold, is of the
$AdS_{2}\times S^{2}$~-type but it does not belong to the  Bertotti-
Robinson class. Moreover, it should be noted that it is the next-to-
leading term of the stress-energy that gives rise to the second term
in the right hand side of (\ref{dfm}), and, consequently the first-
order calculations lead to the Bertotti-Robinson limit. This behavior
is, of course, in concord with our earlier discussion.

To understand this result let us return to the semiclassical Einstein
field equations for the general line element (\ref{ss}) describing
extreme black hole. Putting $f(r_{+}) = f'(r_{+}) =0$  at the
degenerate horizon and taking the trace, one has
\begin{equation}
f''(r_{+}) =\frac{2}{r_{+}^{2}} + 8\pi T_{i}^{i}(r_{+}),
\end{equation}
provided some technical requirements concerning regularity of the line
element are satisfied. Now, let us expand the metric potentials in the
line element as
\begin{equation}
f(r) = P (r-r_{+})^{2} +{\cal O} (r-r_{+})^{3},
\end{equation}
where $P =\frac{1}{2}f''(r_{+}),$
make use of the coordinate transformation
\begin{equation}
r= r_{+} + \Delta  e^{-\tau} \sinh \chi, \hspace{1cm} 
t=\frac{e^{\tau} \coth \chi}{\Delta P} e^{-\psi (r_{+})},
\end{equation}
and, finally, take $\Delta =0$ limit. It can  easily be demonstrated 
that the result of all these operations is the line element (\ref{AdS2}). 
It is simply because
$T_{i}^{(em)i}$ and $T_{i}^{(1)i}(r_{+})$ are traceless whereas 
\begin{equation}
8\pi T_{i}^{(2)i}(r_{+}) = -\frac{11}{4725\pi m^{4} r_{+}^{8}}.
\end{equation}
This explains also why the earlier calculations, which were based on
the first-order approximation to the quantum part of the total stress-
energy tensor yielded the Bertotti-Robinson line element. Indeed, for
the traceless (total) stress-energy tensor one has  $P(r_{+}) =1/r_{+}
^{2}$  and the near-horizon geometry is precisely that of  Bertotti
and Robinson.

\section{Final remarks}

We have calculated explicitly the next-to-leading term of the
approximate renormalized stress-energy tensor of the quantized massive
scalar field in a large mass limit in the Bertotti-Robinson and
Reissner-Nordstr\"om geometry. Additionally, we have calculated the
next-to-next-to-leading term in the Bertotti-Robinson geometry. These
arise from the approximate effective action constructed from the
integrated coincidence limit of the coefficients $a_{4}(x,x')$ and
$a_{5}(x,x').$ It has been shown that although the Bertotti-Robinson
geometry is an exact solution of the semiclassical Einstein field
equations with the source term given by the leading term of the
Schwinger-DeWitt approximation, the addition of the next-to-leading
term makes the resulting equations contradictory  and requires
incorporation of the negative cosmological constant. This behavior is
insensitive to presence of the electric charge.

We have also considered small deformations of the Bertotti-Robinson
manifold leading  to the general $AdS_{2} \times S^{2}$ geometry. Such
geometries (together with the Nariai solutions~\cite{Nariai1,Nariai2})
appear naturally in the closest vicinity of the event horizon of the
extreme black holes. Further, solving the semiclassical Einstein field
equations we have constructed the quantum corrected Reissner-
Nordstr\"om black hole and analyzed its extreme limit. Since the
stress-energy tensor (\ref{sta}-\ref{fin}) improves the approximation
based on the lowest term of the Schwinger-DeWitt action it is expected
that the resulting line element better describes the quantum corrected
geometry than the approximate solutions constructed in Refs.~43,44.
%%%%%
%%% \cite{Taylor,jirinek03b} 
%%%%%
We have shown that the near horizon geometry of the extreme quantum
corrected black hole is, as expected, characterized by
Eqs.~(\ref{GBRr}) and (\ref{hiperbol}).

The calculations carried out so far indicate that the second order
term is a considerable portion of the leading term and as such it must
not be neglected in the serious analyses.

%\bibliographystyle{prsty}
%\bibliography{bacA4,dod}

%%%%%%%%%%
%%%%%%%%%%
%%%%%%%%%%

\end{document}